# Synthesis of New Lithium- and Monoamine-Intercalated Superconductors Li$_x$(C$_n$H$_{2n+3}$N)$_y$Fe$_{1-z}$Se ($n$ = 6, 8, 18) with the Dramatically Expanded Interlayer Spacing

Chika Sakamoto[1], Takashi Noji[1], Kazuki Sato[2], Takayuki Kawamata[1], and Masatsune Kato[1]

[1] *Department of Applied Physics, Graduate School of Engineering, Tohoku University, Sendai 980-8579, Japan*
[2] *Department of Engineering for Future Innovation, National Institute of Technonlogy, Ichinoseki College, Takanashi, Hagisho, Ichinoseki-Shi, Iwate, 021-8511, Japan*

New superconductors, Li$_x$(C$_n$H$_{2n+3}$N)$_y$Fe$_{1-z}$Se ($n$ = 6, 8, 18), have been synthesized via the co-intercalation of linear monoamines together with Li into FeSe. The distance between neighboring Fe layers expands up to 55.7 Å for $n$ = 18, which is much larger than the previous record of 19 Å in the FeSe-based intercalation superconductors. $T_c$ remains saturated at ~ 42 K.

The superconducting transition temperature $T_c$ of FeSe, which has the simplest crystal structure among the iron-based superconductors, is as low as 8 K,[1] but it dramatically increases up to 40 - 45 K by the expansion of the interlayer spacing between neighboring Fe layers, $d$, through the co-intercalation of alkali/alkaline-earth metals and ammonia/organic molecules between the FeSe layers.[2-9] It has been found that $T_c$ increases with increasing $d$ up to ~ 9 Å and then tends to be saturated at 40 - 45 K,[2-9] which indicates that the two-dimensionality of the electronic structure plays an important role in the high-$T_c$ in the FeSe-based superconductors. The largest $d$ = 19 Å is obtained in 2-phenethylamine (2-PEA)-intercalated Li$_x$(C$_8$H$_{11}$N)$_y$Fe$_{1-z}$Se. As shown in Figs. 1(a) and 1(b), two molecules of 2-PEA, which is a monoamine with a nitrogen atom at one end, are intercalated together with two Li$^+$ ions in series perpendicular to the FeSe layers,[9] unlike diamine-intercalated Li$_x$(C$_n$H$_{2n+4}$N$_2$)$_y$Fe$_{1-z}$Se, in which one molecule of diamine with two nitrogen atoms at both ends is intercalated together with two Li$^+$ ions perpendicular to the FeSe layers.[8] Therefore, $d$ > 19 Å can be expected for compounds where a linear monoamine longer than 2-PEA with a length of 8.2 Å is intercalated together with Li.

In this research, we have attempted the synthesis of new Li- and linear-monoamine-intercalated superconductors of Li$_x$(C$_n$H$_{2n+3}$N)$_y$Fe$_{1-z}$Se ($n$ = 6, 8, 18) with $d$ > 19 Å.

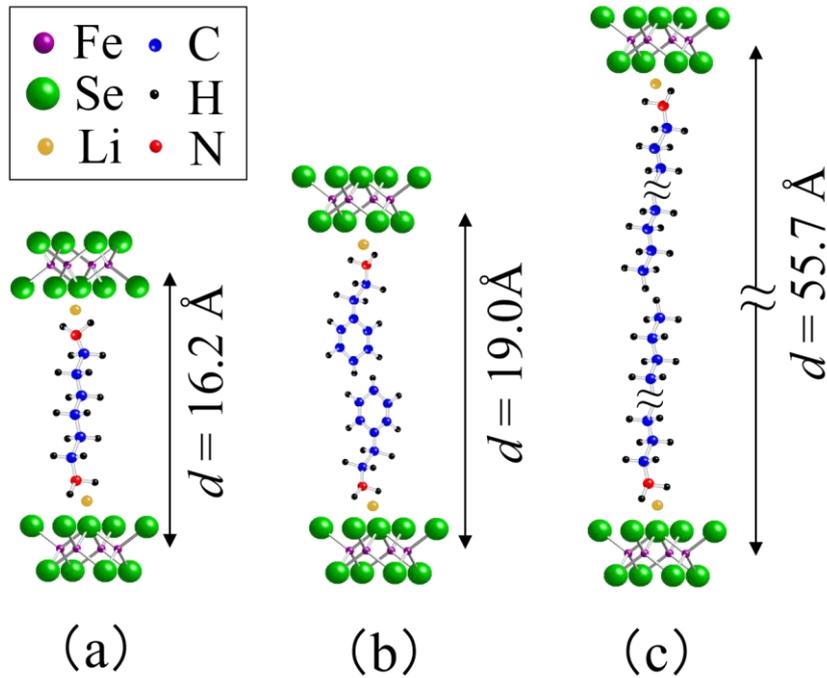

**Fig. 1.** Schematic crystal structures of Li- and amine- intercalated FeSe-based superconductors. (a) $Li_x(HMDA)_yFe_{1-z}Se$ (HMDA: hexamethylenediamine, $C_6H_{16}N_2$, 10.4 Å in length)[8] (b) $Li_x(2\text{-}PEA)_yFe_{1-z}Se$ (2-PEA: 2-phenethylamine, $C_8H_{11}N$, 8.2 Å in length) [9] and (c) $Li_x(ODA)_yFe_{1-z}Se$ (ODA: octadecylamine, $C_{18}H_{39}N$, 24.2 Å in length).

Polycrystalline samples of $Li_x(C_nH_{2n+3}N)_yFe_{1-z}Se$ ($n$ = 6, 8, 18) were prepared via the co-intercalation of three different linear monoamines of hexylamine (HA), octylamine (OA) and octadecylamine (ODA) together with Li into FeSe, respectively. The host samples of FeSe were prepared by the solid-state reaction method as described in Ref. 9. As for the co-intercalation of HA or OA with Li, appropriate amounts of FeSe, lithium, naphthalene and HA or OA were reacted in a beaker at 50°C for 7 - 10 days. As for the co-intercalation of ODA with Li, the reaction did not proceed at 50°C, so it was carried out at 160°C for 7 days by the solvothermal method as described in Ref.10. Typical amounts were 0.3803 g of FeSe, 0.0197 g of lithium, 0.1 g of naphthalene and 10 ml of monoamine. The molar ratio is Li : FeSe = 1 : 1. All processes were performed in an argon-filled glove box. The obtained samples were characterized by the powder x-ray diffraction (XRD) using $CuK_\alpha$ radiation. Measurements of the magnetic susceptibility $\chi$ were performed using a SQUID magnetometer.

Figure 2 shows the powder XRD patterns of the host sample of FeSe and intercalated samples of $Li_x(C_nH_{2n+3}N)_yFe_{1-z}Se$ ($n$ = 6, 8, 18). For the intercalated samples, it is found that new Bragg peaks are observed between $2\theta = 3°$ and 15°. These new peaks are able to be indexed as (00$l$) on the basis of the same anti-PbO-type ($P4/nmm$) structure as FeSe. The $c$-axis lengths of $Li_x(C_nH_{2n+3}N)_yFe_{1-z}Se$ ($n$ = 6, 8, 18) are 23.3, 30.2 and 55.7 Å, respectively, so that the increments of the $d$ value through the co-intercalation are estimated to be 17.8, 24.7 and 50.2 Å, respectively. These values are about twice larger than the lengths of HA (9.2 Å), OA (11.7 Å) and ODA (24.2 Å). This means that two molecules of linear monoamine are intercalated in series perpendicular to the FeSe layers as well as in 2-PEA-intercalated $A_x(C_8H_{11}N)_yFe_{1-z}Se$ ($A$ = Li, Na),[9] though the location of the intercalated Li ion is uncertain, as shown in Fig. 1(c). ODA molecule is much longer than 2-PEA molecule. That is why, the $d$-value of 55.7 Å for $Li_x(C_{18}H_{39}N)_yFe_{1-z}Se$ is greatly larger than the previous record of 19 Å for $Li_x(C_8H_{11}N)_yFe_{1-z}Se$, and is in good agreement with $d$ = 56.1 Å for ODA intercalated $(C_{18}H_{39}N)_yTaS_2$.[11]

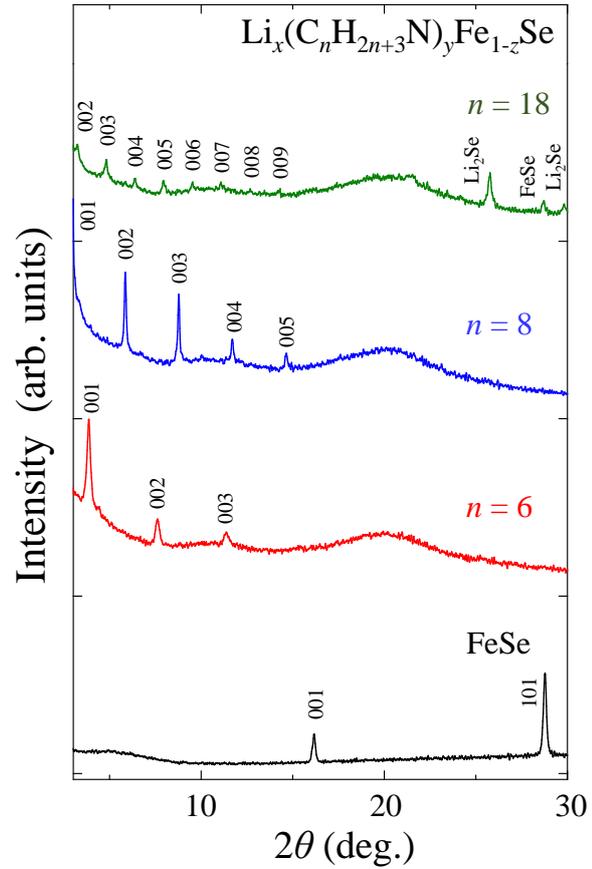

**Fig. 2.** Powder XRD patterns of the host sample of FeSe and the intercalated samples of $Li_x(C_nH_{2n+3}N)_yFe_{1-z}Se$ ($n$ = 6, 8, 18) using CuK$_\alpha$ radiation. The broad peak around $2\theta = 20°$ is due to the airtight sample holder.

Figure 3 shows the temperature dependences of $\chi$ for the intercalated samples of $Li_x(C_nH_{2n+3}N)_yFe_{1-z}Se$ ($n$ = 6, 8, 18). The superconducting transitions are observed for all samples. The shift to the positive direction and the hysteresis above $T_c$ are due to the magnetic impurities generated by the strong reduction in the co-intercalation process, which are also seen in other FeSe-based co-intercalation compounds.[3-9] First of all, as for $Li_x(C_6H_{15}N)_yFe_{1-z}Se$ with $n$ = 6, two superconducting transitions are observed. Considering the results of the powder XRD, the first transition at 43 K is due to $Li_x(C_6H_{15}N)_yFe_{1-z}Se$. On the other hand, the gradual decrease in $\chi$ below ~ 10 K is due to the small amount of the non-intercalated region of FeSe, though the Bragg peaks due to FeSe are not observed in the XRD patterns within our experimental accuracy as shown in Fig. 2. The non-intercalated region of FeSe tends to remain because both Li and long monoamine are hard to diffuse into the inside of FeSe crystals at the reaction temperature of 50°C. Next, as for $Li_x(C_8H_{19}N)_yFe_{1-z}Se$ with $n = 8$, a single superconducting transition is observed around 42 K.

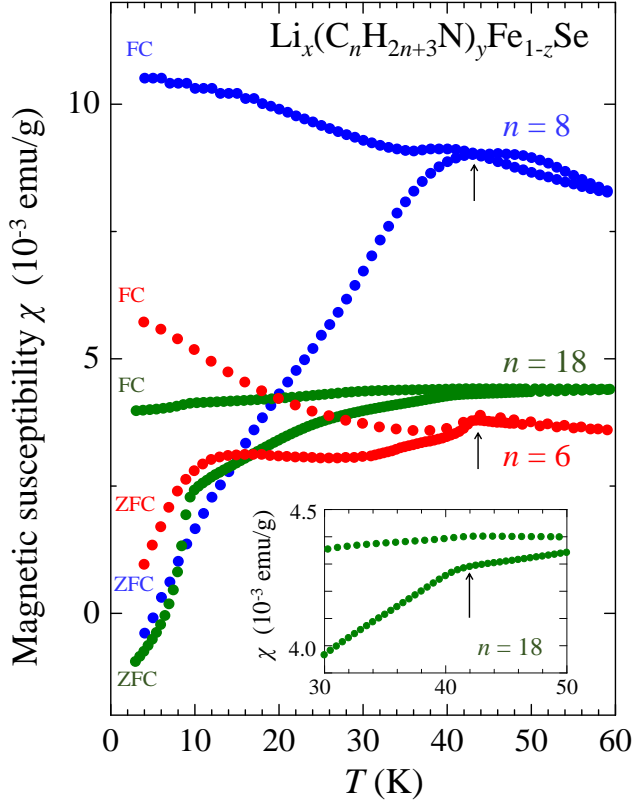

**Fig. 3.** Temperature dependences of the magnetic susceptibility $\chi$ measured in a magnetic field of 10 Oe on zero-field cooling (ZFC) and field cooling (FC) for the intercalated samples of $Li_x(C_nH_{2n+3}N)_yFe_{1-z}Se$. Inset is an enlarged plot for $n = 18$ around $T_c$. Arrows indicate onset temperatures of $T_c$, where $\chi$ starts to deviate from the normal-state value.

The superconducting volume fraction, roughly estimated from the change in $\chi$ between 42 K and 10 K, is largest among the present samples, which is reasonable taking into account the sharp and strong peaks in the powder XRD patterns. Finally, as for $Li_x(C_{18}H_{39}N)_yFe_{1-z}Se$ with $n = 18$, a superconducting transition is detected at 42 K due to $Li_x(C_{18}H_{39}N)_yFe_{1-z}Se$, as shown in the inset in Fig. 2. This small diamagnetic signal may be because the superconducting phase of $Li_x(C_{18}H_{39}N)_yFe_{1-z}Se$ is small in amount and its grain size is small, as shown in the weak and broad XRD peaks. The values of $T_c$, defined at the onset temperature where $\chi$ starts to deviate from the normal-state value, are 43, 42 and 42 K for $Li_x(C_nH_{2n+3}N)_yFe_{1-z}Se$ with $n = $ 6, 8, 18, respectively. It is worth mentioning that $T_c = 42 - 43$ K observed in the present works is not due to the Li-intercalated phase of $Li_xFeSe$ because its $T_c$ is ~ 8 K[12] and that the large interlayer spacing is essential for the increase in $T_c$ from 8 K to 42 - 43 K as previously reported.[2-9] These $T_c$ values are comparable to those for the previous FeSe-based intercalation superconductors with $d = 8 - 19$ Å,[2-9] indicating that $T_c$ remains saturated at 40 - 45 K as the interlayer distance is increased up to 55 Å. This tendency is explained by the theoretical calculation by Guterding *et al.*[13] That is, the enhancement of the two-dimensionality of the electronic structure with the increase of $d$ leads to the increase in $T_c$ and the perfect two-dimensionality is realized at $d$ = 8 - 10 Å, leading to the saturation of $T_c$ at $d \geq 9$ Å. Moreover, the saturation of $T_c$ = 40 - 45 K is consistent with the fact that the single-layer FeSe films, which are regarded as a kind of FeSe based intercalation superconductor with infinite $d$ values, exhibit a superconducting transition at ~ 43 K.[14] Finally, the length of intercalated organic molecules, the interlayer spacing $d$ and $T_c$ in various FeSe-based co-intercalation compounds are summarized in Table I.

**Table I.** Length of intercalated organic molecules, interlayer spacing $d$ and $T_c$ in various FeSe-based co-intercalation compounds.

| intercalated molecules | length of molecules (Å) | $d$ (Å) | $T_c$ (K) | Ref. |
|---|---|---|---|---|
| hexylamine (HA), $C_6H_{15}N$ | 9.2 | 23.3 | 43 | this work |
| octylamine (OA), $C_8H_{19}N$ | 11.7 | 30.2 | 42 | this work |
| octadecylamine (ODA), $C_{18}H_{39}N$ | 24.2 | 55.7 | 42 | this work |
| ethylenediamine (EDA), $C_2H_8N_2$ | 5.4 | 10.4 | 45 | Ref. 5, 6 |
| hexamethylenediamine (HMDA), $C_6H_{16}N_2$ | 10.4 | 16.2 | 41 | Ref. 8 |
| 2-phenethylamine (2-PEA), $C_8H_{11}N$ | 8.2 | 19.0 | 39 | Ref. 9 |

In conclusion, we have successfully synthesized new monoamine-intercalated superconductors of $Li_x(C_nH_{2n+3}N)_yFe_{1-z}Se$ ($n$ = 6, 8, 18). The $d$-value greatly increases up to 55 Å, but $T_c$ remains at ~ 42 K. No further increase in $T_c$ seems to be expected in any FeSe-based superconductors with larger $d$ values. However, the present study has showed that the co-intercalation of linear monoamines with the alkali metals into the layered compounds is useful technique for the investigation of two-dimensional superconductivity.

**Acknowledgements** This work was supported by JSPS KAKENHI (Grant Numbers 19K05241 and 20K14418).